\documentclass[journal,11pt,onecolumn,letterpaper]{IEEEtran}
\usepackage[T1]{fontenc}
\usepackage[latin9]{inputenc}
\usepackage{amsmath}
\usepackage{graphicx}
\usepackage{epsfig,color}
\usepackage{amssymb}
\usepackage{cite}
\usepackage{bm}
\usepackage{enumerate}
\usepackage{setspace}
\usepackage{empheq}
\usepackage{subfigure}
\makeatletter

\usepackage{color}		% Need the color package
\usepackage{epsfig}
\allowdisplaybreaks

\newcommand{\wo}{\bm{h}}
\newcommand{\phio}{\bm{\phi}}
\newcommand{\psio}{\bm{w}}
\newcommand{\uo}{\bm{u}}

\newcommand{\what}{\hat{\bm{h}}}
\newcommand{\whatu}{\underline{\hat{\bm{h}}}}

\newcommand{\dif}{\mathrm{d}_{(\bm{G})}}
\newcommand{\eb}{\bm{\underline{\epsilon}}}
\newcommand{\N}{\mathcal{N}}
\newcommand{\R}{\mathbb{R}}
\newcommand{\Z}{\mathbb{Z}_{\geq 0}}

\newcommand{\J}{\mathcal{J}_n}
\newcommand{\I}{\mathcal{I}_{k,n}}
\newcommand{\Th}{ \Theta_{k,n}(\phio_{k,n})}
\newcommand{\Ths}{ \Theta'_{k,n}(\phio_{k,n})}
\newcommand{\Wb}{{B_{\ell_1}[\psio_n,\rho]}}
\newcommand{\Prb}{P^{(\bm{G})}_{B_{\ell_1}[\psio_n,\rho]}}

\newcommand{\G}{\bm{G}}
\newcommand{\Ghat}{\bm{\underline{G}}}
\newcommand{\levs}{\mathrm{lev}_{\leq0}}

\renewcommand{\QED}{\hfill\IEEEQED}
\title{A Sparsity-Aware Adaptive Algorithm for Distributed Learning}
\author{Symeon Chouvardas \IEEEmembership{Student Member, IEEE,} Konstantinos Slavakis, \IEEEmembership{Member, IEEE,} Yannis Kopsinis \IEEEmembership{Member, IEEE,} and Sergios Theodoridis, \IEEEmembership{Fellow, IEEE}%

\thanks{S. Chouvardas, Y. Kopsinis, and S. Theodoridis are with the Department of Informatics and Telecommunications, University of Athens, Panepistimioupolis,
Ilissia, 157 84 Athens, Greece (e-mail: schouv@di.uoa.gr,kospinis@ieee.org,stheodor@di.uoa.gr),}
\thanks{K. Slavakis is with the Department of Telecommunications Science and
Technology, University of Peloponnese, Tripolis 22100, Greece (e-mail:
slavakis@uop.gr).}}
\begin{document}
 \maketitle
 \abstract In this paper, a sparsity-aware adaptive algorithm for distributed learning in diffusion networks is developed. The algorithm follows the set-theoretic estimation rationale. At each time instance and at each node of the network, a closed convex set, known as property set, is constructed based on the received measurements; this defines the region in which the solution is searched for. In this paper, the property sets take the form of hyperslabs. The goal is to find a point that belongs to the intersection of these hyperslabs. To this end, sparsity encouraging variable metric projections onto the hyperslabs have been adopted. Moreover, sparsity is also imposed by employing variable metric projections onto weighted $\ell_1$ balls. A combine adapt cooperation strategy is adopted. Under some mild assumptions, the scheme enjoys monotonicity, asymptotic optimality and strong convergence to a point that lies in the consensus subspace. Finally, numerical examples verify the validity of the proposed scheme, compared to other algorithms, which have been developed in the context of sparse adaptive learning.
 \section{Introduction}
 Sparsity, i.e., the presence of a few number of non-zero coefficients of a signal/parameter  vector to be estimated, has been  attracting, recently, an overwhelming interest under the Compressed Sensing (CS) framework \cite{donoho2006compressed,candes}.  However, most of the efforts,  so far, have been invested in CS-based signal recovery
techniques, which are appropriate for batch mode operation. Accordingly, the estimation of the signal
parameters can be achieved only after a  fixed number of measurements has been collected and stored. If a new
measurement becomes available, the whole estimation process has to be repeated from scratch. As the
number of measurements increases, the computational burden becomes prohibitive for real time applications.
On the contrary, time-adaptive/online updating succeeds in improving the current estimate
dynamically as new measurements are obtained. Moreover, batch methods are not directly suited for time
varying scenarios, where the parameter vector changes, as time evolves.  Online, learning techniques  overcome the previously mentioned limitations. Online techniques for sparsity-aware learning have recently become the focus of intense research activity,  e.g., \cite{kopsinis,ZA-LMS,sparserls}. 
 
 In this paper, the task of sparsity-aware learning is treated in the context of distributed processing \cite{lopes2008diffusion,mateos2009performance,chouv2010,cattivelli2010}. To be more specific, we consider the typical setup of a distributed network, in which the estimate of the unknown  parameter vector is based on noisy measurements sensed by a number of spatially distributed nodes.
This task can be fulfilled following several approaches, with the centralized solution being one of them. In such a scenario, the nodes transmit the measured information to a central node, called fusion center, which carries out the full amount of computations.  Nevertheless, the existence of a fusion center is not always feasible due to power or geographical constraints. Furthermore, this approach lacks robustness, since if the fusion center is malfunctioning then the network collapses. Hence, in many applications,  a decentralized philosophy has to be followed, in which the nodes themselves take part in the computation task. The most celebrated examples of such networks are: 
\begin{itemize}
\item The incremental, in which each node is able to communicate with only one neighbouring node and, henceforth, the nodes are part of  a cyclic pattern, e.g., \cite{lopes2007incremental,li2010distributed}. This topology requires small bandwidth, albeit it is not robust when a number of  nodes  are malfunctioning, since, when a node fails, the network collapses. 
\item The diffusion, where each node shares information with a subset of nodes. Despite the fact that the diffusion topology requires larger bandwidth, compared to the incremental one, it is robust to cope with node failures, and it's implementation turns out to be easier when large networks are involved \cite{takahashi2010diffusion,cattivelli2010,mateos2009performance,lopes2008diffusion}. 
\end{itemize}
Although there are a few sparsity-aware methods for  batch processing in distributed learning,  e.g., \cite{mota2010,dlasso},  to the best of our knowledge there is no algorithm, yet, capable for time-adaptive/online processing to operate  in diffusion networks. 

The algorithm, to be presented here, handles the requests for sparsity-awareness and operation in diffusion networks, simultaneously. It follows  the set-theoretic estimation rationale \cite{combettes}, that is,
instead of seeking for a (unique) optimum vector, we search for a set of points that
are \textit{in agreement} with the received set of  measurements. To this end, at each time instance, a closed convex set, namely a hyperslab, is defined by the currently received input-output training data pair,  and \textit{any} point that lies within this set is considered to be in agreement with the current measurements. Moreover, following similar philosophy as in \cite{kopsinis}, in order to exploit the a-priori knowledge concerning the sparsity of the unknown vector,  we constrain the search  for a possible solution within sparsity-promoting weighted $\ell_1$ balls.
 The goal becomes that of finding a point that lies in the intersection of the \textit{infinite} number of hyperslabs with the previously mentioned constraint sets; this is successfully solved (see for example \cite{yamada2004adaptive,slavakis2006adaptive,theodoridisp}) by employing a sequence of projections onto the hyperslabs and the weighted $\ell_1$ balls. In the current study, the previous scheme is enchanced by reformulating the projection operators appropriately so as to exploit further the a-priori information with respect to the sparsity of the unknown vector. This can be achieved (see for example \cite{yukawaunified}), by adopting  the \textit{variable metric projections} rationale. As a consequence, the variable metric projections improve the convergence speed, when seeking for a sparse vector, since different weights are assigned at each coefficient of the updated vector, and, through this procedure, small coefficients are forced to  diminish faster. The reasoning of assigning different weights at each coefficient, is also met in the so called proportionate algorithms \cite{PNLMS,benesty2002improved}.

The paper is organized as follows. In section  \ref{probform} the general problem is described and in the next section, adaptive strategies for estimating  sparse signals are provided. In section  
 \ref{distributed}, we shed light on basic concepts regarding adaptive distributed learning and in section \ref{algorithmos} the proposed algorithm, together with its theoretical analysis, is discussed. Finally, in section \ref{experiments} the performance of the proposed algorithm is validated and in the Appendices the theoretical background is discussed, and full proofs of the theorems are given.

\section{Problem Formulation}
\label{probform}
%\onehalfspacing
The set of all real numbers and the set of all non-negative integers are denoted by $\mathbb{R}$ and $\Z$, respectively. Given two integers $j_1,j_2$, with $j_1\leq j_2$, we define $\overline{j_1,j_2}=\lbrace j_1,\ldots,j_2\rbrace$. The stage of discussion will be the Euclidean space $\R^m$, where $m$ is a positive integer. We denote vectors by boldface letters, e.g., $\bm{h}$, and matrices with 	upper-case boldfaced letters. 
Furthermore, we define the weighted inner product  as follows: $\forall \wo_1,\wo_2\in\R^m, \langle \wo_1,\wo_2\rangle_{\bm{V}}:=\wo_1^T\bm{V}\wo_2$, and the weighted norm $\forall \wo\in\R^m,\Vert \wo\Vert_{\bm{V}}=\sqrt{\langle \wo,\wo\rangle_{\bm{V}}}$, where the $m\times m$ matrix, $\bm{V}$, is positive definite, and the notation $ (\cdot)^T$ stands for the transposition operator. The Euclidean norm, i.e., $\Vert{\cdot}\Vert$, is a special case of the previously mentioned norm, and occurs if $\bm{V}=\bm{I}_m$, where $\bm{I}_m$ is the $m\times m$ identity matrix. Moreover, the $2$-norm of a matrix, say $\bm{A}$, is denoted by  $\Vert\bm{A}\Vert$. Given a vector $\wo=[h_1,\ldots,h_m]^T\in\R^m$, the $\ell_1$ norm is defined $\Vert\wo\Vert_{1}:=\sum_{i=1}^m \vert h_i\vert$, and the support set, $\mathrm{supp}(\wo):=\lbrace i\in \overline{1,m}: h_i\neq 0\rbrace$. Finally, the $\ell_0$ "norm" is the cardinality of the support set, i.e.,  $\Vert\wo\Vert_{0}:=\vert\mathrm{supp}(\wo)\vert$, where given a set, say $\mathcal{S}$, the notation $\vert\mathcal{S}\vert$ stands for it's cardinality. 

Consider the problem of estimating an unknown parameter vector $\wo^*\in\R^m$, exploiting measurements 
$(d_n,  \uo_n)_{n\in\Z}\in\R\times\R^{m}$, which are related via the linear system
\begin{equation}
d_n=\uo^T_n\wo^*+v_n,  \ \forall n\in \Z,
\label{linsys}
\end{equation}
where $v_n$ is the noise process. We assume that $\wo^*$  is sparse, i.e., $\Vert\wo^*\Vert_{0} \ll m $, or, in other words, it has a few number of non-zero coefficients. Suppose that a finite number of measurements, say $N$, is available. In that case,  (\ref{linsys}) can be written  as
\begin{equation*}
\bm{d}=\bm{U}\wo^*+\bm{v},
\end{equation*}
where the regression matrix $\bm{U}=[\uo_1,\ldots,\uo_N]^T \in \R^{N\times m}, \ \bm{d}=[d_1,\ldots,d_N]^T\in \R^{N}$, $\bm{v}=[v_1,\ldots,v_N]^T \in\R^{N}$, and $N< m$.
Classical techniques, as for example the celebrated least-squares method, fail to produce a good estimate of the unknown parameters, since the sparsity of $\wo^*$ is not taken into consideration and, consequently, there is no guarantee, for a finite number of measurements, that the  estimate will predict the support, i.e., the set of non-zero components,  and force the rest to become zero.  This results at an increased  misadjustment between the true and the estimated values, \cite{baraniuk2007compressive}.
Nevertheless,  one can resort to a
sparsity promoting technique, namely Least Absolute Shrinkage and Selection Operator (Lasso), and overstep the previously mentioned problem. Analytically, the Lasso estimator promotes sparsity, by solving the following optimization task
\begin{equation*}
\hat{\wo}=\mathrm{argmin}_{\Vert \wo \Vert_{1}\leq \delta} \Vert \bm{d}-\bm{U}\wo \Vert^2,
%\label{lasso}
\end{equation*}
where the term $\Vert \bm{d}-\bm{U}\wo \Vert$ accounts for the error residual in the estimation process, and  the $\ell_1$ norm promotes sparsity by shrinking small coefficient values towards  zero, e.g., \cite{Bruck}. 
Most of the emphasis in solving the Lasso problem has been given on batch techniques, see, e.g., \cite{Male}. However, such techniques are inappropriate for online learning, where data arrive sequentially and/or the environment is not stationary but it undergoes changes as time evolves.
 
\section{Sparsity-aware adaptive algorithms}\label{sparsityaware}
Although sparsity promoting adaptive algorithms have drawn the attention of the signal processing community for many years, see, e.g., \cite{benesty2002improved,PNLMS}, it is only recently that the topic is being treated in a more theoretically sound framework, within the spirit of $\ell_1$ regularization, e.g.,  \cite{kopsinis, ZA-LMS, sparserls, murakami, Mileounis}. The a-priori information concerning the underlying sparsity is provided via a constraint built around the $\ell_1$ norm.  Providing this a-priori information,  the convergence rate is improved significantly, and the associated error floor in the steady state is reduced, as well. 

As it is  often the case, most of these efforts evolve along the three main axes in adaptive filtering. One is along  the gradient descend rationale, as this is represented in the adaptive learning by the LMS \cite{ZA-LMS, Mileounis}. The other direction follows Newton-type arguments, as represented by the RLS \cite{sparserls}. The other route is more recent and builds upon  recent extensions of the classical Projections Onto Convex Sets (POCS) theory, which allow for  applications in the online time-adaptive setting, e.g., \cite{yamada2004adaptive,slavakis2006adaptive,slavakis2011,theodoridisp}.  Our new algorithm belongs to this last category and it exploits its potential to allow for convex constraints to be efficiently incorporated within the algorithmic flow. 

\subsection{Set-theoretic estimation approach and variable metric projections}
In this paper, the set-theoretic estimation rationale e.g., \cite{diniz2008adaptive, theodoridisp,combettes}, will be adopted. The philosophy behind this family of algorithms is that instead of adopting a loss function to be optimized, in order to obtain an estimate of the unknown target parameter vector, one obtains an estimate that lies in the intersection of an infinite  number of convex sets. Each one of the (convex) property sets,  is constructed using the information that is provided by the respective measurement pair $(d_n,\bm{u}_n)$, and basically defines, in turn, a region where the unknown vector lies with a high probability, based on the received information and the assumed nature of the noise source.  We say that such a convex set is  ``in agreement'' with the received measurement pair. Moreover, in the presence of convex constraints, each of them  defines a convex region and the solution is searched in the intersection of all the involved sets, those associated with the measurements as well as those with the constraints. 

The strategy used in order to achieve the previously mentioned goal of finding a point that lies in the intersection of the infinite number of convex sets  was presented in \cite{yamada2004adaptive}. This algorithmic scheme can be seen as a generalization of the POCS theory \cite{pocs1,combettes,pocs2}. The difference lies in the fact that in the classical POCS theory,  a finite number of convex sets is involved.  On the contrary, in its adaptive version, an infinite number of sets are involved.  In the adaptive setting, the task of identifying a point in the intersection of convex sets, is accomplished by projecting in parallel, the currently available estimate over the $q$ most recently ``received'' sets. This provides the new estimate. If constraints are present, e.g., \cite{slavakis2006adaptive}, further projections are performed one for each of the constraint sets (the definition of the projection is given in Appendix A). Under some mild assumptions, the estimates converge to a point that lies in the intersection of all the involved convex sets. 

It has been pointed out (see, for example, \cite{yukawaunified}), that the sparsity-related a-priori knowledge can be  ``embedded''  in the projection operators to the benefit of the algorithm's performance. To this end,  the notion of the {\it variable metric projection}  is introduced.  The result of a variable metric projection  of a certain vector, onto a closed convex set (see also Appendix A), is determined by: a) a positive definite matrix, which defines the induced inner product, b) the convex set, onto which the projection takes place, c)  the vector, which is projected. The difference with the classical standard metric projections (Appendix A) is that in the latter the  matrix, that defines the weighted inner product,  is the simplest case of a positive definite matrix, i.e., the identity one. As it will become clear later on, for a properly chosen  matrix, which is time-dependent and it is constructed via the current estimate at each time instance, the variable metric projection pushes small coefficients to diminish faster. In other words, by employing at each time instance a different inner product in our Euclidean space, we manage to change the topology of the space in order to favour sparse solution vectors.

\begin{figure}
\centering
 \includegraphics[scale=0.6]{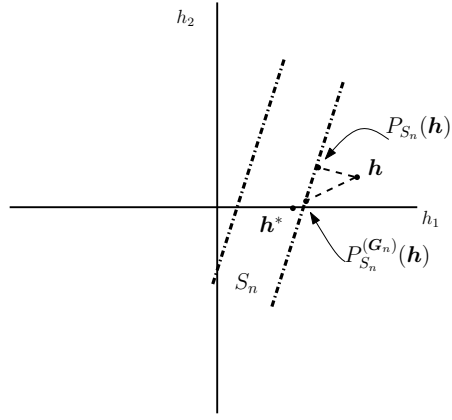}
\label{projfig}
%}
\caption{ Illustration of a hyperslab, the standard metric projection of a vector $\wo$ onto it, denoted by $P_{S_n}(\wo)$, and the variable metric projection onto it.}
\label{ipnlmsfig}
\end{figure}

In the current paper, the adopted property sets,  in which one seeks for a candidate solution, take the form of hyperslabs, i.e.,
\begin{equation}
S_{n}:=\lbrace\wo\in\R^m: \vert d_n-\uo_n^T\wo\vert \leq\epsilon\rbrace,
\end{equation}
where $\epsilon\geq 0$ is a user-defined parameter.    The parameter $\epsilon$ serves as a threshold and it takes into consideration the noise, as well as possible inaccuracies in the adopted model. In this setting, any point that lies within this hyperslab is in agreement with the current measurement. The choice of a hyperslab, in order to define the property sets, is in line with criteria that have been proposed in the context of the robust statistics rationale, e.g., \cite{theodoridisp,huberbook}.  
The { variable metric projection} onto the respective hyperslabs is  defined as \cite{yukawaset}:
 \begin{equation}
\forall \bm{h}\in\R^m, \quad P^{(\G_n)}_{{S}_{n}}({{\bm{h}}})=
\bm{h}+\beta_{n}\G^{-1}_{n}{\bm{u}}_{n},
\label{vprojhyper}
\end{equation}
where
\begin{equation*}
{\beta}_{n}=
\begin{cases}
\dfrac{d_{n} - {\bm{u}}_{n}^T{\bm{h}}   +\epsilon}{\Vert{\bm{u}}_{n} \Vert_{\G^{-1}_{n}}^2}, & \mathrm{if} \ d_n-{\bm{u}}_{n}^T{\bm{h}} < -\epsilon, \\
0, & \mathrm{if} \ \vert d_n-{\bm{u}}_{n}^T{\bm{h}} \vert \leq \epsilon, \\
\dfrac{d_n-{\bm{u}}_{n}^T{\bm{h}} -\epsilon}{\Vert{\bm{u}}_{n} \Vert_{\G^{-1}_{n}}^2}, & \mathrm{if} \ d_n-{\bm{u}}_{n}^T{\bm{h}} > \epsilon.
\end{cases}
\end{equation*}
Note that if  $\G_n=\bm{I}_m$, then (\ref{vprojhyper}) is the  standard metric projection onto a hyperslab. The positive definite diagonal matrix $\G^{-1}_n$ is constructed following similar philosophy as in \cite{benesty2002improved,yukawaunified}. The $i$-th coefficient of its diagonal equals to $g_{i,n}^{-1}=\frac{1-\alpha}{m}+\alpha\frac{\vert h_i^{(n)}\vert}{\Vert\wo_n\Vert_{1}}$, where $\alpha\in[0,1)$ is a parameter, that determines to which extend the sparsity level of the unknown vector will be taken into consideration, and $h_i^{(n)}$ denotes the $i$-th component of $\wo_n$. Now, in order to grasp the reasoning of the variable metric projections, consider the ideal situation, in which $\G^{-1}_n$ is generated by the unknown vector $\wo^*$. It is easy to verify that $g_{i,n}^{-1}>g_{i',n}^{-1}, \mbox{if}~i\in\mathrm{supp}(\wo^*),~\mbox{and}~i'\notin \mathrm{supp}(\wo^*)$. Hence, employing the variable metric projection, the amplitude of each coefficient of the vector used to construct $\G^{-1}_n$ determines the weight that will be assigned to the corresponding coefficient of the second term of the right hand side in (\ref{vprojhyper}). That is, components with larger  magnitude are weighted heavier than those of lower magnitude. 
Loosely speaking, the variable metric projections accelerate the convergence speed when tracking a sparse vector, due to the fact that the  procedure of assigning different weights makes the coefficients of the estimates with small amplitude, to diminish faster. The geometric implication of it is that the projection is made to ``lean'' towards the direction of the more significant components of the currently available estimate. Obviously,
since $\wo^*$ is unknown, in order to assign the previously mentioned
weights, we rely on the available estimate of it, i.e., $\wo_n$, at
each time instance.   These concepts are depicted in Fig. \ref{ipnlmsfig}.

%Summarizing, the variable metric projection, takes into consideration the amplitude of the coefficients of the vector to be projected, and assigns to the components of the projected vector a proper weight.
%Finally,  the parameter $\alpha$ determines how much the sparsity of the unknown vector will be taken into consideration. 

\textbf{Remark 1:}
The variable metric projections rationale is in line with the so called \textit{proportionate} algorithms \cite{benesty2002improved,PNLMS,yukawa2009krylov}. At the heart of these algorithms lies the fact that at every time instance different weights are assigned to the coordinates of the vector, which produces the next estimate.\QED %Besides, as it will become %clear later on, the algorithmic scheme that will be presented here is a generalization of the Improved Proporionate Normalized %Least Mean Squares (IPNLMS) algorithm \cite{benesty2002improved}.
% Nevertheless, given a certain vector $\wo$, the result of the projection onto the hyperslab is determined by $\G^{-1}_n\uo_n$. So, in practice, if through the adaptation process, we have obtained an estimate $\wo_n$  for which $\vert h_i^{(n)}\vert < \vert h_i'^{(n)}\vert,  i\in\mathrm{supp}(\wo^*),i'\notin \mathrm{supp}(\wo^*)$ then the variable metric projection 

 %So, the difference between these two projection techniques, lies in the fact that in the variable metric projection, a different weight, occurring by the current estimate, is assigned at each coefficient of the, whereas in the Euclidean projection, the coefficients have the same weights. First of all, notice that the Euclidean projection onto this set occurs if $\G^{-1}_{n}=\bm{I}_m$. Obviously, given a certain vector  $\wo$, the projection of it onto the hyperslab is determined by the second term of the right hand side of (\ref{vprojhyper}). Hence, if in the vector $\G^{-1}_n\uo_n$ small weighs are assigned $\forall i\notin\mathrm{supp}(\wo^*)$ and larger weights are assigned $\forall i\in\mathrm{supp}(\wo^*)$, this would result in an attraction

%\subsection{Imposing sparsity by projecting onto weighted $\ell_1$ balls}

\begin{figure}
\centering
%\subfigure[]{
%\includegraphics[scale=0.5]{Differentiable.pstex}
%\includegraphics[scale=0.6]{hyperslab}
%\scalebox{0.5}{\input{Differentiable.pstex_t}} 
%\caption{blah blah.}
%\label{hyperslab}
%}
%\subfigure[]{
 \includegraphics[scale=0.6]{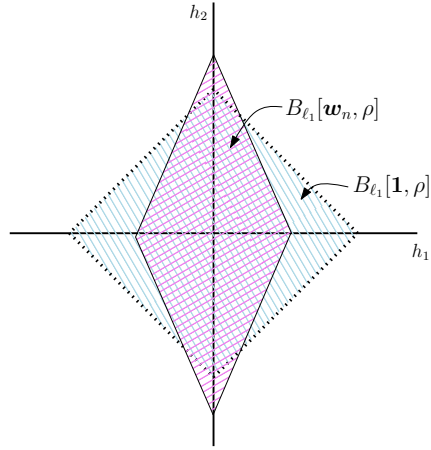}
\label{wl1}
%}
\caption{ Illustration of a weighted $\ell_1$ ball (solid line magenta) and an unweighted $\ell_1$ ball (dashed line blue).}
\label{costfunctions}
\end{figure}

As a second step, in order to exploit the sparsity of the unknown vector, sparsity promoting  constraints, which  take the form of  $\ell_1$ balls, are employed. In order to enhance convergence speed, the notion of the weighted $\ell_1$ ball will be adopted \cite{candes}. A sparsity-aware adaptive scheme, based on set-theoretic estimation arguments, in which the constraints are weighted $\ell_1$ balls, was presented in \cite{kopsinis}.  Given a vector of weights $\psio_n=[w_1^{(n)},\ldots,w_m^{(n)}]^T$, where $w_i^{(n)}>0,\forall i=1,\ldots,m,$ and a positive radius, $\rho$, the weighted $\ell_1$ ball is defined as:
$\Wb:=\lbrace\wo\in\R^m:\sum_{i=1}^m w_i^{(n)}\vert h_i\vert \leq\rho\rbrace$. Notice, that the classical $\ell_1$ ball occurs if $\psio_n=\bm{1}$, where $\bm{1}\in\R^m$ is the vector of ones. The projection onto $\Wb$, is given in \cite[Theorem 1]{kopsinis}, and the geometry of these sets is illustrated in Fig. \ref{costfunctions}.

 It was shown, that the estimates of the algorithm proposed in \cite{kopsinis} converge asymptotically to a point, that lies arbitrarily close to the intersection of the hyperslabs  with the weighted $\ell_1$ balls, with the possible exception of a finite number of outliers. In this paper, 
a generalized version of the algorithm presented in \cite{kopsinis}, will be developed in the next section. 

\textbf{Remark 2:}
The weighted $\ell_1$ ball is determined by the vector of weights, and the radius. Strategies of constructing the weights have been proposed in \cite{kopsinis,candes}. More specifically, $w_i^{(n)}=1/(\vert h_i^{(n)}\vert+\tilde{\epsilon}_n),\ i=1,\ldots,m$ , where $\tilde{\epsilon}_n$ is a sequence of positive numbers, used in order to avoid divisions by zero. It has been shown, e.g., \cite{kopsinis}, that by choosing the weights according to the previously mentioned strategy,  a necessary condition  that guarantees  convergence of the algorithm to the unknown parameter is to set $\rho\geq\Vert \wo^*\Vert_{0}$, since then it holds that $\wo^*\in\Wb$.
 %If the previously mentioned parameter is large, then the algorithm exhibits a fast convergence speed, albeit the steady state error floor is high. On the contrary, if  $\tilde{\epsilon}_n$ is small, then the convergence speed is degraded but the error floor remains low. Hence, a possible strategy, which is adopted in \cite{kopsinis}, is to begin with a large initial value which will be decreasing as time steps increase. Analytically, $\tilde{\epsilon}_n=\tilde{\epsilon}'+1/(n+1),\forall n\in\Z$, for a sufficiently small value $\tilde{\epsilon}'$. Such a choice does not disturb the convergence speed and leads to a low steady state error floor. 
\QED

Here we should note that in  \cite{kopsinis}, standard metric projections onto the hyperslabs and the weighted $\ell_1$ balls take place. However, as it will become clear in Appendix C, since we use variable metric projections onto the hyperslabs, the  induced inner product, which will be used in the analysis of the algorithm, is time varying and it is determined by the matrix $\G_n$. This fact forces us to employ variable metric projections onto the respective $\ell_1$ balls too.

\textbf{Claim 1:}
Recall the  definition of the diagonal matrix $\G_n$. The variable metric projection onto $\Wb$ is given by $P^{(\bm{G}_n)}_{B_{\ell_1}[\psio_n,\rho]}=\G^{-\frac{1}{2}}_n P_{B_{\ell_1}[\G^{-\frac{1}{2}}_n\psio_n,\rho]}\G^{\frac{1}{2}}_n$.

\textbf{Proof:}
The proof is given in Appendix B.\QED
\section{Adaptive distributed learning}
\label{distributed}
\begin{figure}
  \begin{center}
     \includegraphics[scale=0.5]{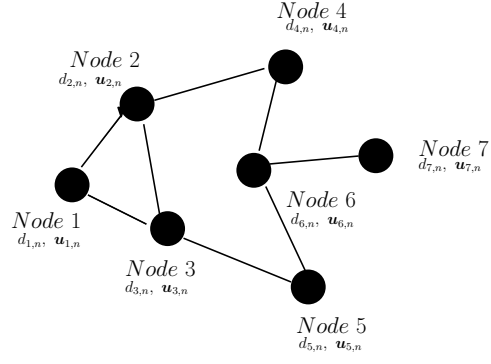}
  \end{center}
  \caption{ Illustration of a diffusion network with $K=7$ nodes.}
  \label{diffusion}
\end{figure}

We now come to the main point of this paper.  Our task is to estimate the sparse, unknown parameter vector $\wo^*\in\R^m$, exploiting measurements collected at the $K$ nodes of a network obeying the diffusion topology. 
An example of such a network is illustrated in Fig. \ref{diffusion}. The node set is denoted by $\N=\lbrace 1,\ldots,K \rbrace$ and we assume that each node is able to communicate, i.e., to exchange information, with a subset of $\N$, namely $\N_k,\ k=1,\ldots,K$. This set, hereafter, will be called the \textit{neighbourhood of $k$}. Moreover, each node has access to the measurement pair $\left(d_{k,n},\uo_{k,n}\right)_{n\in\Z}, \ k \in \N$, where $\uo_{k,n}\in \R^m$ and $d_{k,n}\in \R$, and the measurements are related according to $d_{k,n}=\uo_{k,n}^T\wo^*+v_{k,n}$, where $v_{k,n}$ stands for the additive noise at each node.
%
% \begin{equation}
% d_k(n)=\uo_{k,n}^T\wo^*+v_k(n), \forall n\in \Z, \ k\in \N,
% \label{system}
% \end{equation}
 In a nutshell, what differentiates the adaptive distributed learning from the classical adaptive counterpart is the fact that in the former case, each node, besides the locally received measurement pair,  also exploits information received by its neighboring nodes.  For a fixed node, say $k$, and at every time instance, this extra  information comprises the estimates of the unknown vector, which have been obtained, at the previous time instance, from the nodes with which communication is possible, i.e., $\forall l\in\N_k$. The use of this extra information results in a faster convergence speed, as well as a lower steady state error floor, compared to the case where the measurement pair is solely used, e.g., \cite{cattivelli2010, lopes2008diffusion}. One more objective,  which makes the  exchange of the estimates crucial, is that the distributed ``nature'' of our problem imposes the need for \textit{consensus}; this means that the nodes will have to converge to the same estimate. It has been shown, that this information exchange can lead asymptotically to consensus \cite{mateos2009distributed, mateos2009performance, cavalcante2009adaptive, chouv2010}.

Depending on the way with which the estimates are exploited, the following cooperation strategies have been proposed:
\begin{itemize}
\item Combine Adapt, in which, at every node, the estimates from the neighborhood are fused under a certain protocol, and then the aggregate is put into the adaptation step \cite{lopes2008diffusion, chouv2010,chouvkrylov}. 
\item Adapt Combine, where prior to the combination step, comes the adaptation one \cite{cavalcante2009adaptive, cattivelli2010}.
\item Consensus based, where the computations are made in parallel and there is no clear distinction between the combine and the adapt step \cite{mateos2009performance,mateos2009distributed}.
\end{itemize}
Now, let us shed light on the combination of the estimates coming from the neighbourhood of each node. Recall the previous discussion; an arbitrary node, $k$, is able to communicate with every node that belongs to $\N_k$. We assume that the following hold true: $k\in\N_k,\forall k \in \N$ and $l \in \N_k \Leftrightarrow k \in \N_l, \ \forall k,l\in \N$. Moreover, we consider that the network is strongly connected, i.e., there is a, possibly multihop, path that connects every two nodes of the network. These assumptions are very common in adaptive distributed learning (see for example \cite{lopes2008diffusion, mateos2009performance}). As stated earlier, the estimates, received from the neighborhood, are fused under a certain protocol. The most common strategy is to take a linear combination of the estimates. To be more specific, we define the combination coefficients, for which we have that $c_{k,l}(n)>0$, if $l\in \N_k$, $c_{k,l}(n)=0$, if $l\notin \N_k$ and $\sum_{l\in\N_k}c_{k,l}(n)=1$. From the previous definition, it can be readily seen that every node assigns a \textit{weight} to each one of the estimates which are received from the neighborhood. Two well known examples of combination coefficients are: the Metropolis rule, where 
\begin{equation*}
c_{k,l}(n)=
\begin{cases}
\frac{1}{\mathrm{max}\left\lbrace\vert\N_k\vert,\vert\N_l\vert\right\rbrace}, & \text{if $l\in\N_k$ and $l\neq k$},\\
1 - \sum_{l\in\N_k \setminus k}c_{k,l}(n), & \text{if $l=k$},\\
0,& \mathrm{otherwise},
\end{cases}
\end{equation*}
and the uniform rule, in which the coefficients are defined as
\begin{equation*}
c_{k,l}(n)=
\begin{cases}
\frac{1}{\vert\N_k\vert}, & \text{if $l\in\N_k$},\\
0,& \mathrm{otherwise}.
\end{cases}
\end{equation*}
Collecting all the coefficients for a network, we define the combination matrix $\bm{C}_{n}$, in which the $k,l$-th component is $c_{k,l}(n)$. This matrix gives us information about the network's topology, as if the $k,l$-th entry is equal to zero, this implies that the nodes $k, \ l$ are not connected. The opposite also holds true, since a positive coefficient implies that the nodes are connected.  Finally, we define the $Km\times Km$ consensus matrix, $\bm{P}_{n}=\bm{C}_{n}\otimes \bm{I}_m$, where the symbol $\otimes$ stands for the Kronecker product. Some very useful properties of this matrix are \cite{cavalcante2009adaptive}: 
\begin{enumerate}[1)]
\item $\Vert \bm{P}_{n} \Vert = 1$.
\item Any consensus matrix $\bm{P}_n$ can be decomposed as \begin{equation*}\bm{P}_{n}=\bm{X}_{n} + \bm{B} \bm{B}^T,\label{decomp}\end{equation*} 
where $\bm{B}=[\bm{b}_{1},\ldots,\bm{b}_{m}]$ is an $Km\times m$ matrix,
and $\bm{b}_{k}=\dfrac{(\bm{1}_{K}\otimes\bm{e}_{k})}{\sqrt{K}}$,
 $\bm{e}_{k}$ is a $m\times1$ vector of zeros except the
$k$-th entry, which is one and $\bm{X}_{n}$ is an $Km\times Km$ matrix for which it holds that $\left\Vert \bm{X}_n \right\Vert <1$.
\item $\bm{P}_{n} \underline{\breve{\wo}}=\underline{\breve{\wo}}, \forall \underline{\breve{\wo}} \in \mathcal{O}:= \lbrace\underline{\bm{h}} \in \mathbb{R}^{Km}: \ \underline{{\wo}}= [ \bm{h}^T,\ldots,\bm{h}^T ]^T, \  \bm{h} \in \mathbb{R}^m \rbrace$. The subspace $\mathcal{O}$ is the so called consensus subspace of dimension $m$, and $\bm{b}_k,\ k=1,\ldots,m$, constitute a basis for this set. Hence, the orthogonal projection of a vector,  $\underline{\bm{h}}$, onto this linear subspace is given by $P_{\mathcal{O}}(\underline{\bm{h}}):=\bm{B}\bm{B}^T\underline{\bm{h}}, \ \forall \underline{\bm{h}}\in\mathbb{R}^{Km}$.
\end{enumerate}
\section{Proposed Algorithmic Scheme}
\label{algorithmos}
\begin{figure}
     \includegraphics[scale=0.8]{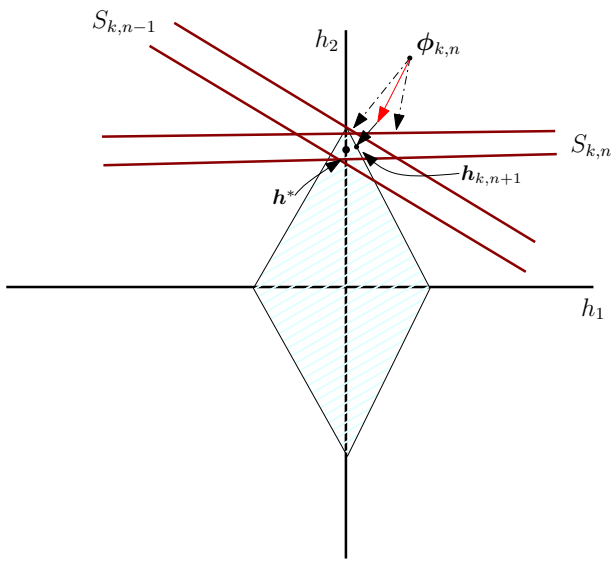}
  \caption{Geometrical interpretation of the algorithm. The number of hyperslabs onto which $\phio_{k,n}$ is projected, using variable metric projections, is $q=2$. The result of these two projections, which are illustrated by the dash dotted black line, is combined (red line) and the result is projected (solid black line) onto the sparsity promoting weighted $\ell_1$ ball, in order to produce the next estimate.}
  \label{diffffusion}
\end{figure}

The goal is to bring together the sparsity promoting "tools", which where discussed in  section \ref{sparsityaware}, and to reformulate  them in a distributed fashion by adopting the combine adapt strategy, which was presented in the previous section. 
The main steps of the algorithm, for each node $k$, at  time instance $n$, in order to produce the next estimate,  can be summarized as follows:

\textbf{Algorithm:}
\begin{enumerate}
\item The estimates from the neighbourhood are received and combined with respect to the adopted combination strategy, in order to produce $\phio_{k,n}=\sum_{l\in\N_k}c_{k,l}(n)\wo_{l,n}, \forall k \in \N$.
\item  Exploiting the newly received measurements $d_{k,n},\uo_{k,n}$ the following hyperslab is defined: $S_{k,n}=\lbrace\wo\in\R^m:\vert d_{k,n}-\uo_{k,n}^T\wo\vert\leq\epsilon_k\rbrace$, where the parameter $\epsilon_k$ is allowed to  vary from node to node.
 The aggregate $\phio_{k,n}$ is projected, using variable metric projections, onto the $q$ most recent hyperslabs, constructed locally, and a convex combination of them  is computed. Analytically,  the sliding window  $\J:=\overline{\max\lbrace 0,n-q+1 \rbrace, n}$ is defined, and it determines the hyperslabs that will be considered at time instance $n$. Given the set of weights 
$\forall j\in\J$, $\omega_{k,j}$, where $\sum_{j\in\J}\omega_{k,j}=1,\forall k\in \N$, the convex combination of the projections onto the hyperslabs, i.e., $\sum_{j\in\J}\omega_{k,j}P_{S_{k,j}}^{(\G_n)}(\phio_{k,n})$ is computed.  The effect of projecting onto a $q>1$ number of hyperslabs is to speed up convergence \cite{kopsinis}.
\item The result of the previous step is projected onto the sparsity constraint set, i.e., the weighted $\ell_1$ ball. 
\end{enumerate}
The previous steps can be encoded in the following mathematical formula:

\begin{equation}
\wo_{k,n+1}=P^{(\bm{G}_n)}_{B_{\ell_1}[\psio_n,\rho]}\left(\phio_{k,n}+\mu_{k,n}\left(\sum_{j\in\J}\omega_{k,j}P^{(\G_n)}_{S_{k,j}}(\phio_{k,n})-\phio_{k,n}\right)\right),
\label{algok}
\end{equation}
where  $\mu_{k,n}\in(0,2\mathcal{M}_{k,n})$, and  
\begin{equation}
\mathcal{M}_{k,n}:=
\begin{cases}
\dfrac{\sum_{j\in\J}\omega_{k,j} \left\Vert P^{(\G_n)}_{S_{k,j}}(\phio_{k,n})-\phio_{k,n} \right\Vert^2}{\left\Vert  \sum_{j\in\J}\omega_{k,j} P^{(\G_n)}_{S_j}(\phio_{k,n})-\phio_{k,n} \right\Vert^2}, & \mathrm{if} \  \sum_{j\in\J}\omega_{k,j} P^{(\G_n)}_{S_{k,j}}(\phio_{k,n})\neq \phio_{k,n} \\
1, &\mathrm{otherwise}.
\end{cases}
\label{Mu}
\end{equation}
The algorithm has an elegant geometrical interpretation which can be seen in Fig. \ref{diffffusion}.
It turns out that the weighted $\ell_1$ ball, as well as $\G_n$ have to be the same for every node of the network, which yields that this information cannot be constructed locally. This fact, as it will be established in the theoretical analysis of the algorithm, is essential in order to guarantee consensus. Hence, a reasonable strategy, which will be adopted here, is to construct $\psio_n$ and $\G_n$, using the methodology described in section \ref{sparsityaware}, via $\wo_{k_{opt},n}$, where $k_{opt}$ is the node with the smallest noise variance. It is obvious that this requires knowledge, in every node, of $\wo_{k_{opt},n}$, something that is in general infeasible. However, it is not essential to update the parameters at every time instance; instead, $\psio_n$ and $\G_n$ can be updated at every, say $n'\geq 1$, time instances, where $n'$ are the time steps required for $\wo_{k_{opt},n}$ to be distributed over the network. Experiments regarding the robustness of the proposed algorithm with respect to $n'$ are given in the Numerical Examples section. Moreover, as it will become clear in the
Numerical Examples section, it turns out that the algorithm is robust
in cases where the knowledge of the less noisy node is not available,
and/or in cases where the assumption that these quantities must
be common to all nodes is violated and each node uses the locally
available values.

Regarding the complexity of the algorithm, it has been shown in \cite{kopsinis}, that if standard metric projections take place, then the complexity of the respective algorithm is $O(qm)$ coming from the projection operators and $O(m\mathrm{log}_2 m)$ occurring from the projection onto the weighted $\ell_1$ ball. If we employ the variable metric projections, at each node, it is obvious that the term $\G^{-1}_n\uo_{k,j}, \ j\in\J$ has to be computed, and this adds $qm$ multiplication operations. %Nevertheless, exploiting the fact that $\uo_{k,n}:=[u_{k,n},\ldots,u_{k,n-m+1}]^T$ the following strategy is adopted.
%\begin{enumerate}
%\item Compute $\G^{-1}_n\uo_{k,n}$ for which we need $m$ multiplications.
%\item Since $\uo_{k,n}$ and $\uo_{k,n-1}$ differ only at one coefficient 
%compute $g_m^{(n)}u_{k,n-m}$ and retrieve the rest coefficients from the previous step.
%\end{enumerate}
%It can be readily verified that with the previous procedure, the variable metric projections add %$m+q-1$ multiplications in the overall complexity of the algorithm. 

\textbf{Remark 3:} The algorithm presented in \cite{kopsinis} is a special case of the scheme in (\ref{algok}), if $K=1$ and $\G_n=\bm{I}_m$. The same also holds for the IPNLMS \cite{benesty2002improved} if we let $K=1$, $q=1$, $\epsilon_k=0$ and $\Prb=I$, where $I$ stands for the identity operator.\QED

As it will be verified in Appendix C, the algorithm in (\ref{algok}) enjoys monotonicity, asymptotic optimality and strong convergence to a point that lies in the consensus subspace. The  assumptions under which the previous hold are the following. 

\textbf{Assumptions.}
\begin{enumerate}[(a)]
\item Define $\forall n \in\Z, \ \Omega_n=\Wb\cap\left(\bigcap_{j\in\J}\bigcap_{k\in\N}S_{k,j}\right)$. Assume that there exists $n_0\in\Z$, such that $\Omega:=\bigcap_{n\geq n_0}\Omega_n\neq\emptyset$.
\item There exists $n_1\in\Z$, such  that $\bm{G}_{n}=\bm{G}_{n_1}=:\bm{G},\forall n\geq n_1$. In other words, the update of the  matrix $\bm{G}_{n}$ pauses after a finite number of iterations\footnote{Notice that the matrix $\bm{G}_{n}$ is constructed via $\wo_{k_{opt},n}$, hence  $\forall n\geq n_1$, the variable metric projections is determined by $\wo_{k_{opt},n_1}$. In practice, for sufficiently large $n_1$, the algorithm has converged and the fact that $\G_n$ is not updated does not affect the performance of the algorithm.}.
\item Assume a sufficiently small $\varepsilon_1$, such that $\forall k\in\N,\frac{\mu_{k,n}}{\mathcal{M}_{k,n}}\in [\varepsilon_1,2-\varepsilon_1]$.
\item  Assume $\forall k\in\N \ \tilde{\omega}_k:=\inf\lbrace\omega_{k,j}:j\in\J, n\in \Z\rbrace>0$.
\item
Define $\mathfrak{C}:=\bm{\Omega} \cap \mathcal{O}$, where the cartesian product space $\bm{\Omega}:=\underbrace{\Omega\times\ldots\times\Omega}_{K}$.  We assume that $\mathrm{ri}_{\mathcal{O}}\bm{\Omega}\neq\emptyset$, where this term stands for the relative interior of $\mathfrak{C}$ with respect to $\mathcal{O}$ (see Appendix A).
\end{enumerate}

\textbf{Theorem 1:}
Under the previous assumptions, the following hold:
\begin{enumerate}[(1)]
\item \textbf{Monotonicity.} Under assumptions (a), (b), (c), it holds that   $\forall n\geq z_0, \ \forall \whatu\in\mathfrak{C}, \Vert\bm{\underline{h}}_{n+1}-\whatu\Vert_{\Ghat} \leq \Vert\bm{\underline{h}}_{n}-\whatu\Vert_{\Ghat}$, where $z_0:=\max\lbrace n_0,n_1\rbrace$, $\Ghat$ is the $Km \times Km$ block-diagonal matrix, with definition $\Ghat:=\mathrm{diag}\underbrace{\lbrace\bm{G},\ldots,\bm{G}\rbrace}_K$, and $\bm{\underline{h}}_{n}=[\wo^T_{1,n},\ldots,\wo^T_{K,n}]^T\in\R^{Km},\forall n\in\Z$. 
\item \textbf{Asymptotic Optimality.} If  assumptions (a), (b), (c), (d) hold true then $ \lim_{n\rightarrow\infty}\max\lbrace\mathrm{d}(\wo_{k,n+1},S_{k,j}):j\in\J\rbrace=0, \forall k\in\N,$ where $\mathrm{d}(\cdot,S_{k,j})$ denotes the distance of $\wo_{k,n+1}$ from $S_{k,j}$ (see Appendix A). The previous implies that the distance of the estimates from the respective hyperslabs will tend  asymptotically to zero. 
\item \textbf{Asymptotic Consensus.} Consider that assumptions  (a), (b), (c), (d) hold. Then $\lim_{n\rightarrow\infty}\Vert\wo_{k,n}-\wo_{l,n}\Vert=0, \ \forall k,l\in\N$. 
\item \textbf{Strong Convergence.} Under assumptions (a), (b), (c), (d), (e), it holds that  $\lim_{n\rightarrow\infty}\bm{\underline{h}}_n=\hat{\underline{\bm{h}}}_*,\hat{\underline{\bm{h}}}_*\in\mathcal{O}$. So, the estimates for the whole network, converge to a point that lies in the consensus subspace.
\end{enumerate}
\textbf{Proof:} The proof is given in Appendix C.\QED
%For each node, fix a $q\in\Ze$, which is the number of hyperslab onto which we project in a step and define the following sliding window $\J:=\overline{\max\lbrace 0,n-q+1 \rbrace, n}$. Furthermore
%\begin{figure}
%\centering
% \includegraphics[scale=0.6]{algorithmos}
%\label{algorithm}
%\caption{ Illustration of a weighted $\ell_1$ ball (magenta) and an unweighted $\ell_1$ ball (blue).}
%\end{figure}

\section{Numerical Examples}\label{experiments}
In this section, the performance of the proposed algorithm is validated within the system identification framework. Due to the fact that the online algorithmic schemes, proposed in the literature, cover non-distributed learning scenarios, in the first experiment we compare the proposed algorithm against others in the context of a  non-distributed system identification task. This essentially allow us to evaluate the variable metric projections scheme, since this is one of the contributions of this paper. More specifically, we compare the proposed algorithm with the  Adaptive Projection based algorithm using Weighted $\ell_1$ Balls (APWL1) \cite{kopsinis}, with the Online Cyclic Coordinate Descent  Time Weighted Lasso (OCCD-TWL), the Online Cyclic Coordinate Descent Time
and Norm Weighted LASSO (OCCD-TNWL), both proposed in  \cite{sparserls}, and with the LMS-based, Sparse Adaptive Orthogonal Matching Pursuit  (Spadomp) \cite{Mileounis}. The unknown vector is of dimension $m=512$ and the number of non-zero coefficients, equals to $20$. Moreover, the input samples $\uo_n=[u_n,\ldots,u_{n-m+1}]^T$ are drawn from a Gaussian distribution, with zero mean and standard deviation equal to $1$. The noise process   is Gaussian with variance equal to $\sigma^2=0.01$. Finally, the adopted performance metric, which will be used, is the average Mean Square Deviation (MSD), given by $\mathrm{MSD}(n)=1/K \sum_{k=1}^K\Vert \wo_{k,n}-\wo^* \Vert^2$, and the curves occur from an averaging of $100$ realizations for smoothing purposes.

\begin{figure}
\center
     \includegraphics[scale=0.5]{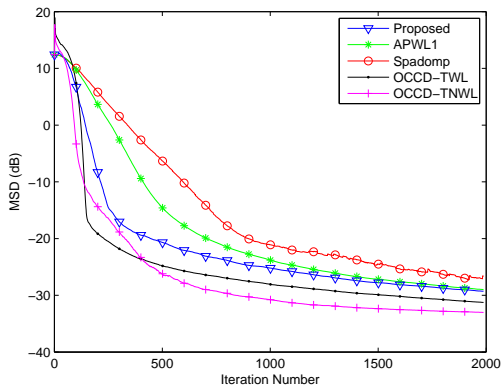}
  \caption{MSD for the experiment 1.}
  \label{experiment1}
\end{figure}
In the projection-based algorithms, i.e., the proposed and the APWL1, the number of hyperslabs used per time update equals to $q=55$, the width of the hyperslabs equals to $\epsilon=1.3\times\sigma$,   and the step-size equals to $\mu_n=0.2\times\mathcal{M}_n$, where $\mathcal{M}_n$ is given in (\ref{Mu}), and the node subscript is omitted. Moreover, for the  weights we choose $\omega_{n}=1/q$. These choices are not necessarily optimal, albeit they lead to a good trade-off between the convergence speed and the steady state error floor.  The radius of the weighted $\ell_1$ ball equals to $\rho=\Vert\wo^*\Vert_{0}$, the weights are constructed according to the discussion in section \ref{sparsityaware}, and $\tilde{\epsilon}_n=10^{-2}$. Furthermore,  the weighting matrix $\G_n$ is defined according to the strategy presented in section \ref{sparsityaware}. Regarding the parameter $\alpha$, we observed that a value close to $1$ leads to a fast convergence speed but it increases the steady state error floor, and vice versa. So, at the beginning of the adaptation, we choose $\alpha=0.99$ and at every $250$ time instances, we set $\alpha=\alpha/2$.  Finally, $\psio_n$ and $\G_n$ are updated at every time instance, i.e., $n'=1$.
In the OCCD-TWL and the OCCD-TNWL, the regularization parameter is chosen to be $\lambda_{\mathrm{TWL}}=\sqrt{2\sigma^2 n \mathrm{log}m},\lambda_{\mathrm{TNWL}}=\sqrt{2\sigma^2 n^{4/3} \mathrm{log}m} $, respectively, as adviced in \cite{sparserls}. The step size, adopted in the Spadomp,  equals to $0.2$, due to the fact that this choice gives similar steady state error floor with the projection-based algorithms\footnote{Extensive experiments have shown that a choice of a smaller step-size, results in a slower convergence speed, without significant improvement in the steady state error floor.}. The forgetting factor of OCCD-TWN, OCCD-TNWL and Spadomp equals to 1 since, in the specific example, the system under consideration does not change with time.  
\begin{figure}
\center
     \includegraphics[scale=0.5]{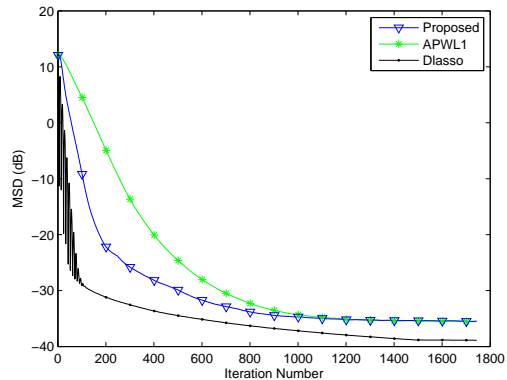}
  \caption{MSD for the experiment 2.}
  \label{experiment2}
\end{figure}
From Fig. \ref{experiment1}, it can be seen that the proposed algorithm exhibits faster convergence speed compared to  the APWL1 to the common error floor. Moreover, the proposed algorithm outperforms the Spadomp, since it converges faster and the steady state error floor is slightly better. We should point out, that the complexity of the Spadomp is $O(m)$, which implies that for the previously mentioned choice of $q$, the proposed algorithm is of larger complexity.  
Compared to the OCCD-TWL, we observe that its performance  is slightly better, compared to the proposed one, albeit the complexity of the algorithm is $O(m^2)$. Finally, the OCCD-TNWL outerforms  the rest of the algorithms, at the expense of a higher complexity, which is approximately twice  that of OCCD-TWL.

In the second experiment, we consider a network consisted of $K=10$ nodes, in which the nodes are tasked to estimate an unknown parameter $\wo^*$ of dimension $m=256$. The number of non-zero coefficients, of the unknown parameter equals to $20$ and each node has access to the measurements $(d_{k,n},\uo_{k,n})$, where the regressors are defined as in the previous experiment. The variance of the noise  at each node is $\sigma_k^2=0.01\varsigma_k$, where $\varsigma_k\in[0.5,1]$, following the uniform distribution.
We compare the proposed algorithm with the distributed APWL1, i.e., the proposed if we let $\G_n=\bm{I}_m$, and the distributed Lasso (Dlasso)  \cite{dlasso}. The Dlasso is a batch algorithm, which implies that the  data have to be available prior to start the processing.  So, here we assume that at every time instance, in which a new pair of data samples becomes available, the algorithm is re-initialized so as to solve a new optimization problem.
For the projection-based algorithms, $q=20$ and the rest of the parameters are chosen as in the previous experiment. Moreover, the combiners $c_{k,l}(n)$ are chosen with respect to the Metropolis rule. Finally, the regularization parameter in the Dlasso is set via the distributed cross-validation procedure, which is proposed in \cite{dlasso}. From Fig. \ref{experiment2} we observe that the Dlasso outperforms the projection-based algorithms and that the proposed algorithm converges faster than APWL1. %Through experiments we observed that a larger choice of $q$, would lead the convergence speed of the proposed algorithm close to the one of the Dlasso. 
However, for $q=20$, the complexity of the proposed algorithm is significantly lower than that of the Dlasso.  Dlasso, at every time instance, requires the inversion of a $m\times m$ matrix. %is of order $O(20m)$, whereas for the Dlasso is $O(m^2)$.  
\begin{figure}
\center
     \includegraphics[scale=0.5]{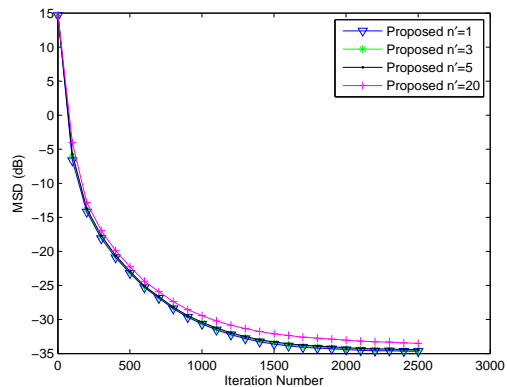}
  \caption{MSD for the experiment 3.}
  \label{experiment3}
\end{figure}

In the third experiment, we study the sensitivity of the proposed algorithm to choice of the parameter $n'$, i.e., the frequency at which $\psio_n$ and $\G_n$ are updated. To this end, the parameters are the same as in the previous experiment, but we set different values to $n'$. Fig. \ref{experiment3} illustrates that the algorithm is relatively insensitive to the frequency of the updates, since even in the case where $n'=20$ the algorithm exhibits fast convergence speed. This is important, since the robustness of the proposed scheme to choice of the parameter $n'$ makes it suitable to be adopted in distributed learning.% in distributed networks.

\begin{figure}
\center
     \includegraphics[scale=0.5]{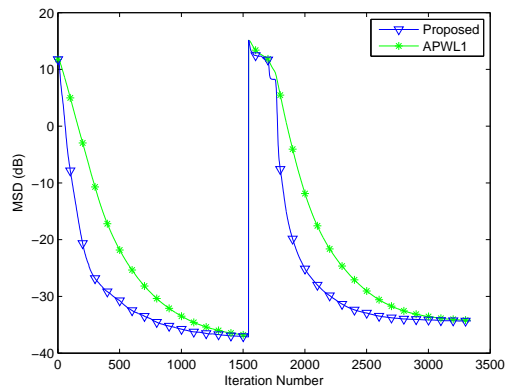}
  \caption{MSD for the experiment 4.}
  \label{experiment4}
\end{figure}

In the fourth experiment, we validate the performance of the algorithm in a non-stationary environment. It is by now well established that a fast convergence speed does not necessarily imply a good tracking ability \cite{haykin}. More specifically, we consider that a sudden change in the unknown parameter takes place. So, until $\wo^*$ changes, the parameters remain the same as in the second experiment, and after the sudden change, we have that $\Vert\wo^*\Vert_{0}=15$. The radius of the weighted $\ell_1$ ball is set equal to $23$, due to the fact that through experiments we observed relative insensitiveness  of the performance of the proposed algorithmic scheme to choices of $\rho$, as long as it remains larger   than $\Vert \wo^*\Vert_0$. Furthermore, we assume the algorithm is able to monitor sudden changes of the orbit $(\wo_{k,n})_{n\in\Z}$, in order to reset the value of $\alpha$ when the channel changes. To be more specific,  we reset the value of $\alpha$, if the ratio $\Vert \wo_{k,n+1}-\wo_{k,n}\Vert/\Vert\wo_{k,n}-\wo_{k,n-1}\Vert, \ \forall k\in\N,$    is greater than a threshold, which is chosen, here, to be equal to $10$. This strategy is adopted since we observed that  if the algorithm has converged, the previously mentioned ratio  takes values close to $1$, whereas if an abrupt change takes place in the unknown parameter, then the value of the ratio  increases significantly. From Fig. \ref{experiment4}, it can be observed that both the projection-based algorithms enjoy good tracking ability, when a sudden change occurs. Moreover, as in the previous experiments, the proposed algorithm converges faster than the APWL1 to a similar error floor.

\begin{figure}
\center
     \includegraphics[scale=0.5]{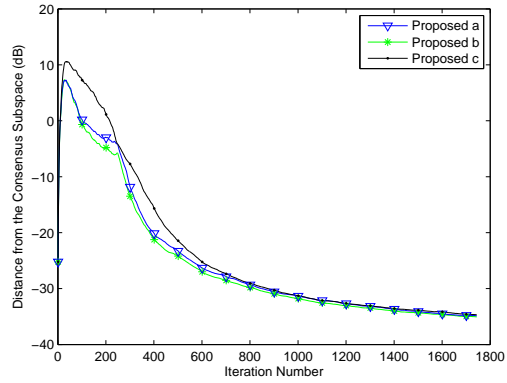}
  \caption{Squared distance from the consensus subspace, for experiment 5.}
  \label{exp5}
\end{figure}

Finally, in the fifth experiment, we study the robustness of the
proposed scheme, with respect to adopting different strategies in order
to construct $\psio_n$ and $\G_n$. To this end, we consider the following
strategies: a) the previously mentioned quantities are constructed using
the node with the smallest noise variance (Proposed a), b) $\psio_n$ and $\G_n$ are generated via the node with the largest variance (Proposed b)
and c) $\psio_n$ and $\G_n$ are constructed locally at every node (Proposed
c). Obviously, the latter one violates  the theoretical assumption
of having common weights to all nodes. In order to verify whether
the nodes reach consensus, we plot the squared distance of $\underline{\wo}_n$ from the consensus subspace, i.e., $\Vert\underline{\wo}_n-P_{\mathcal{O}}(\underline{\wo}_n)\Vert^2$. As in the previous experiments, the curves occurs from an averaging of $100$ independent experiments. From Fig. \ref{exp5}, it can be readily seen that the distance of $\underline{\wo}_n$ from the consensus subspace, is decreasing as time steps increase. It is interesting, that even in the Proposed c where the assumption, under which asymptotic consensus is achieved, is violated the estimates for the whole network  tend asymptotically to the consensus subspace.
\section{Conclusions}
A sparsity-aware adaptive algorithm for distributed learning has been proposed. The algorithm  builds upon set-theoretic estimation arguments. In order to exploit the sparsity of the unknown vector, variable metric projections onto the hyperslabs within which we seek for a possible solution take place. Moreover, extra projections onto sparsity promoting weighted $\ell_1$ balls are employed in order to enhance further the performance of the proposed scheme. Full convergence analysis has been derived. Numerical examples, within the system identification task, demonstrate  the comparative performance of the proposed algorithm against other recently published algorithms. 

\appendices
\section{Basic Concepts Of Convex Analysis}
The stage of discussion will be $\R^m$ and the induced inner product, given a positive definite $m\times m$ matrix $\bm{V}$, is $\langle\wo_1,\wo_2\rangle_{\bm{V}}=\wo_1^T\bm{V}\wo_2$. A set $\mathcal{C} \subseteq \mathbb{R}^m $, for which it holds that $\forall  \bm{h}_1,\bm{h}_2 \in \mathcal{C} \ \mathrm{and} \  \forall t \in [0,  1], \ t \bm{h}_1 + (1-t) \wo_2 \in \mathcal{C}$, is called convex. Moreover, a function $\Theta: \ \mathbb{R}^m \rightarrow \mathbb{R}$ will be called convex if $\forall  \wo_1,\wo_2 \in \mathbb{R}^m $ and $\forall  t \in [0, 1]$ the inequality $\Theta(t\wo_1 + (1-t)\wo_2) \leq t \Theta(\wo_1) + (1-t)\Theta(\wo_2)$ is satisfied. Finally,  the subdifferential of $\Theta$ at an arbitrary point, $\wo$, is defined as the set of all subgradients of $\Theta$ at $\wo$ (\hspace{-0.1pt}\cite{bertsekas2003convex,hiriart1993convex}), i.e.,
\begin{equation*}
\partial_{(\bm{V})}\Theta(\wo) := \lbrace\bm{s} \in \mathbb{R}^m: \ \Theta(\wo)+\langle\bm{x}-\wo,\bm{s} \rangle_{\bm{V}} \leq \Theta(\bm{x}), \ \forall \bm{x} \in \mathbb{R}^m \rbrace. 
%\label{eqsub}
\end{equation*} 

 The distance of an arbitrary point $\bm{h}$ from a closed non-empty convex set $\mathcal{C}$, with respect to  $\bm{V}$, is given by the distance function 
 \begin{align*} \mathrm{d}^{(\bm{V})}( \cdot , \mathcal{C} )&: \mathbb{R}^m \rightarrow [0,  +\infty) \nonumber \\ &: \bm{h} \mapsto \inf{ \lbrace \Vert \bm{h} - { \bm{x} } \Vert_{\bm{V}} : { \bm{x} } \in \mathcal{C} \rbrace }, 
 \end{align*}
and if we  let $\bm{V}$ be the identity matrix, the Euclidean distance is given.
This function is continuous, convex, nonnegative  and is equal to zero for every point that lies in $\mathcal{C}$ \cite{hiriart1993convex}.  Moreover, the projection mapping, $P^{(\bm{V})}_{\mathcal{C}}$ onto $\mathcal{C}$,  is defined as $P^{(\bm{V})}_{\mathcal{C}}(\wo):=\mathrm{argmin}_{\bm{x}\in\mathcal{C}}\Vert\wo-\bm{x}\Vert_{\bm{V}}$, and as in the distance function, if  $\bm{V}=\bm{I}_m$ the standard metric projection is obtained.

Finally, the relative interior of a nonempty set, $\mathcal{C}$, with respect to another one, $\mathcal{S}$,  is defined as
\begin{equation*}
\text{ri}_{\mathcal{S}}(\mathcal{C})=\lbrace\bm{h} \in \mathcal{C}:\exists \varepsilon_0 >0 \ \text{with} \ \emptyset \neq (B_{(\bm{h}_0,  \varepsilon_0)}\cap \mathcal{S})\subset \mathcal{C}\rbrace, 
\end{equation*}
where $B_{(\bm{h}_0,  \varepsilon_0)}$ is the open ball with definition $B_{(\bm{h}_0,  \varepsilon_0)}:=\lbrace\wo\in\R^m:\Vert\wo-\wo_0\Vert<\varepsilon_0\rbrace$ (see for example \cite{boyd2004convex}), with center $\wo_0$ and radius equal to $\varepsilon_0$.
\section{Variable Metric Projection onto the Weighted $\ell_1$ Ball}
The variable metric projection of $\wo$, onto $\Wb$, is given by
\begin{align*}
&\min_{\bm{x}\in\Wb} \Vert \wo-\bm{x}\Vert^2_{\G_n}\\
&\mathrm{s.t.} \sum_{i=1}^m w^{(n)}_i \vert{x}_i\vert\leq \rho,
\end{align*}
where $\bm{x}:=[x_1,\ldots,x_m]^T$. However, $\Vert \wo-\bm{x}\Vert^2_{\G_n}=\Vert \G^{\frac{1}{2}}_n \left(\wo-\bm{x}\right)\Vert^2=\Vert \G^{\frac{1}{2}}_n\wo-\bm{\xi}\Vert^2$, where  $\bm{\xi}:=\G^{\frac{1}{2}}_n\bm{x}$. Moreover, $\bm{x}=\G^{-\frac{1}{2}}_n\bm{\xi} \Leftrightarrow x_i={\sqrt{g_{i,n}^{-1}}}\xi_i, i=1,\ldots,m $, where $\xi_i$ are the coefficients of $\bm{\xi}$. From the previous, it holds that $ \sum_{i=1}^m w^{(n)}_i \vert{x}_i\vert= \sum_{i=1}^m {\sqrt{g_{i,n}^{-1}}}{w}^{(n)}_i \vert \xi_i\vert$. Hence the initial optimization problem, is equivalent to 
\begin{align*}
&\min_{\bm{\xi}}  \Vert \G^{\frac{1}{2}}_n \wo-\bm{\xi}\Vert^2\\
&\mathrm{s.t.} \sum_{i=1}^m {\sqrt{g_{i,n}^{-1}}}{w}^{(n)}_i \vert\xi_i\vert\leq \rho.
\end{align*}
The solution of the previous optimization, is the standard metric projection of $\G^{\frac{1}{2}}_n\wo$ onto $B_{\ell_1}[\G^{-\frac{1}{2}}_n{\bm{w}}_n,\rho]$ and it can be found in \cite{kopsinis}. So, from the previous $\bm{\xi}_{\mathrm{opt}}=P_{B_{\ell_1}[\G^{-\frac{1}{2}}_n{\bm{w}}_n,\rho]}(\G^{\frac{1}{2}}_n\wo)\Leftrightarrow P^{(\G_n)}_{B_{\ell_1}[\G^{-\frac{1}{2}}_n{\bm{w}}_n,\rho]}(\wo)=\G^{-\frac{1}{2}}_n P_{B_{\ell_1}[\G^{-\frac{1}{2}}_n{\bm{w}}_n,\rho]}(\G^{\frac{1}{2}}_n\wo)$.
\section{Proof of Theorem 1}
\subsection{Monotonicity}
\textbf{Lemma 1:} Define the following non-negative loss functions, $\forall k\in\N$:
\begin{equation}
\forall n\in\Z, \ \forall \wo\in\R^m, \quad \Theta_{k,n}(\wo):=
\begin{cases}
\sum_{j\in\I}\frac{\omega_{k,j}\dif(\phio_{k,n},S_{k,j})}{L_{k,n}}\dif(\wo,S_{k,j}),& \mathrm{if} \ \I\neq\emptyset,\\
0, & \mathrm{if} \ \I=\emptyset,
\end{cases}
\end{equation}
where $\I:=\lbrace j\in\J: \phio_{k,n}\notin S_{k,j}\rbrace$ and $L_{k,n}:=\sum_{j\in\J}\omega_{k,j}\dif(\phio_{k,n},S_{k,j})$. Then (\ref{algok}) is equivalent to\footnote{The time dependence on $\G_n$ is omitted for simplicity in notation.} 
\begin{equation}
\forall n\in\Z, \ \forall k\in\N, \quad \wo_{k,n+1}=
\begin{cases}
P^{(\bm{G})}_{B_{\ell_1}[\psio_n,\rho]}\left( \phio_{k,n}-\lambda_{k,n}\dfrac{\Theta_{k,n}(\phio_{k,n})}{\Vert\Theta_{k,n}'(\phio_{k,n})\Vert_{\bm{G}}^2} \Theta_{k,n}'(\phio_{k,n}) \right),& \mathrm{if} \ \I\neq\emptyset,\\
P^{(\bm{G})}_{B_{\ell_1}[\psio_n,\rho]}\left( \phio_{k,n}\right), &\mathrm{if} \ \I=\emptyset,
\end{cases}
\label{theta}
\end{equation}
where $\Theta_{k,n}'(\phio_{k,n})$ is the subgradient of the function and $\lambda_{k,n}\in(0,2)$.

\textbf{Proof:} 
First of all, notice that if $\I\neq\emptyset$, then there exists $j_0\in\J$ such that $\phio_{k,n}\notin S_{k,j_0} \Leftrightarrow \dif(\phio_{k,n},S_{k,j_0})>0$. Hence, $L_{k,n}\geq \omega_{k,j_0}\dif(\phio_{k,n},S_{k,j_0})>0$, which implies that the denominator in (\ref{theta}) is positive and the cost function is well defined. Now, a subgradient of the distance function, i.e., $\dif(\cdot,S_{k,j})$, is the following \cite{yukawaquadtratic}:
 \begin{equation}
 \mathrm{d}'_{(\bm{G})}(\wo,S_{k,j})=
 \begin{cases}
 \dfrac{\wo-P^{(\bm{G})}_{S_{k,j}}(\wo)}{\dif(\wo,S_{k,j})}, & \mathrm{if} \ \wo\notin S_{k,j}   \\
 \bm{0}, &\mathrm{otherwise}. 
 \end{cases}
 \label{dsub}
 \end{equation}
Recalling basic properties of the subdifferential (see for example \cite{hiriart1993convex}), we have that 
 \begin{equation}
\partial\Theta_{k,n}(\bm{h})=
 \begin{cases}
\sum_{j\in\I}\frac{\omega_{k,j}\dif(\phio_{k,n},S_{k,j})}{L_{k,n}}\partial\dif(\wo,S_{k,j}),& \mathrm{if} \ \I\neq\emptyset,\\
\lbrace\bm{0}\rbrace, & \mathrm{if} \ \I=\emptyset. 
 \end{cases}
 \label{parttheta}
 \end{equation}
 So,  combining (\ref{dsub}), (\ref{parttheta}) and  if $\I\neq\emptyset$ we have
   \begin{align}
   \Ths &=\sum_{j\in\I}\dfrac{\omega_{k,j}\dif(\phio_{k,n},S_{k,j})}{L_{k,n}} \dfrac{\phio_{k,n}-P^{(\bm{G})}_{S_{k,j}}(\phio_{k,n})}{\dif(\phio_{k,n},S_{k,j})} \nonumber \\
   &=\dfrac{1}{L_{k,n}} \sum_{j\in\I}{\omega_{k,j} \left(\phio_{k,n}-P^{(\bm{G})}_{S_{k,j}}(\phio_{k,n})\right)} \nonumber \\
   &=\dfrac{1}{L_{k,n}} \sum_{j\in\J}{\omega_{k,j} \left(\phio_{k,n}-P^{(\bm{G})}_{S_{k,j}}(\phio_{k,n})\right)}.
\label{Ths}   
   \end{align}
Nevertheless, since   $\I\neq\emptyset$, then there exists $j_0\in\J$ such that $\phio_{k,n}\notin S_{k,j_0} \Leftrightarrow P^{(\bm{G})}_{S_{k,j}}(\phio_{k,n})\neq \phio_{k,n}$. So, if $\I\neq\emptyset$ then $\Ths\neq\bm{0}$.
   Following similar steps as in \cite{kopsinis}, it can be proved that $\forall n\geq z_0, \forall j\in\J,  \ \forall k \in \N, \ \Ths=\bm{0}\Leftrightarrow \phio_{k,n}= \sum_{j\in\J} \omega_{k,j} P^{(\bm{G})}_{S_{k,j}}(\phio_{k,n})$. From this fact, if we define $\mu_{k,n}:=\mathcal{M}_{k,n}\lambda_{k,n}$, and if we substitute (\ref{Ths})
in (\ref{theta}) the lemma is proved.\QED

\textbf{Claim 2:} It holds that $\Vert \bm{P}\underline{\bm{h}}-\breve{\underline{\bm{h}}}\Vert_{\Ghat} \leq \Vert \underline{\bm{h}}-\breve{\underline{\bm{h}}} \Vert_{\Ghat}$, $\forall \breve{\underline{\bm{h}}}\in\mathcal{O},
\forall \underline{\bm{h}}\in\R^{Km}$, where $\bm{P}$ is a $Km\times Km$ consensus matrix with $\Vert\bm{P}\Vert=1$.

\textbf{Proof:}
From the definition of $\Vert\cdot\Vert_{\Ghat}$, it can be readily seen that $\Vert \bm{P}\underline{\bm{h}}-\breve{\underline{\bm{h}}}\Vert_{\Ghat}=\Vert {{\Ghat}}^{\frac{1}{2}}\left( \bm{P}\underline{\bm{h}}-\breve{\underline{\bm{h}}}\right)\Vert=\Vert {{\Ghat}}^{\frac{1}{2}} \bm{P} \left( \underline{\bm{h}}-\breve{\underline{\bm{h}}}\right)\Vert$, where this holds since $\breve{\underline{\bm{h}}}\in\mathcal{O}$. 
Moreover, $\underline{\bm{h}}=\left[\begin{matrix}
\bm{h}_1 \\ 
\vdots \\
\bm{h}_K\\ 
\end{matrix}\right],\wo_k\in\R^m,k\in\N$ and $\breve{\underline{\bm{h}}}\in\mathcal{O}\Leftrightarrow \breve{\underline{\bm{h}}}=\left[\begin{matrix}
\breve{\bm{h}} \\ 
\vdots \\
\breve{\bm{h}}\\ 
\end{matrix}\right],\breve{\bm{h}}\in\R^m$.
Recalling the definition of the consensus matrix, with coefficients $c_{k,l},k,l\in\N$, we have the following
\begin{align}
\Vert {{{\Ghat}}}^{\frac{1}{2}} \bm{P} \left( \underline{\bm{h}}-\breve{\underline{\bm{h}}}\right) \Vert &= \left\Vert \left[\begin{matrix}
\bm{G}^{\frac{1}{2}}\sum_{l\in\N_1}c_{1,l} \left(\bm{h}_l - \breve{\bm{h}}\right) \\ 
\vdots \\
\bm{G}^{\frac{1}{2}}\sum_{l\in\N_K}c_{K,l} \left(\bm{h}_l - \breve{\bm{h}}\right) \\ 
\end{matrix}\right] \right\Vert \nonumber\\
&= \left\Vert \left[\begin{matrix}
\sum_{l\in\N_1}c_{1,l} \bm{G}^{\frac{1}{2}}\left(\bm{h}_l - \breve{\bm{h}}\right) \\ 
\vdots \\
\sum_{l\in\N_K}c_{K,l} \bm{G}^{\frac{1}{2}}\left(\bm{h}_l - \breve{\bm{h}}\right) \\ 
\end{matrix}\right] \right\Vert \nonumber \\
&= \left\Vert \bm{P} \left[\begin{matrix}
 \bm{G}^{\frac{1}{2}}\left(\bm{h}_1 - \breve{\bm{h}}\right) \\ 
\vdots \\
 \bm{G}^{\frac{1}{2}}\left(\bm{h}_K - \breve{\bm{h}}\right) \\ 
\end{matrix}\right] \right\Vert \leq \Vert\bm{P}\Vert \left\Vert \left[\begin{matrix}
 \bm{G}^{\frac{1}{2}}\left(\bm{h}_1 - \breve{\bm{h}}\right) \\ 
\vdots \\
 \bm{G}^{\frac{1}{2}}\left(\bm{h}_K - \breve{\bm{h}}\right) \\ 
\end{matrix}\right]  \right\Vert \nonumber \\
&=  \left\Vert  \underline{\wo}-\breve{\underline{\bm{h}}} \right\Vert_{{\Ghat}}
\label{gmonot}
\end{align}
From (\ref{gmonot}), our claim is proved.\QED
%However, since $\bm{G}$ is a diagonal matrix, then $\bm{\widehat{G}}$ is also a diagonal matrix.  

First of all, given a convex function $\Theta:\R^m\rightarrow\R$, with non-empty level set, where the level set is defined $\levs\Theta:=\lbrace\wo\in \R^m:\Theta(\wo)\leq 0\rbrace$,  let us define the subgradient projection mapping, as follows $T_{\Theta}^{(\G)}:\R^m\rightarrow \R^m$ \cite{yukawaquadtratic}:
\begin{equation*}
T_{\Theta}^{(\G)}(\wo):=
\begin{cases}
\wo-\dfrac{\Theta(\wo)}{\Vert\Theta'(\wo)\Vert_{\G}^2}\Theta'(\wo), & \wo\notin\levs\Theta \\
\wo, & \wo\in\levs\Theta,
\end{cases}
%\label{spm}
\end{equation*}
where $\Theta'(\wo)$ is any subgradient of $\Theta$, at $\wo$.
Similarly, we define the relaxed subgradient projection mapping, $T_{\Theta,\lambda}^{(\G)}(\wo):=I+\lambda(T_{\Theta}^{(\G)}(\wo)-I),\lambda\in(0,2)$, where $I$ is the identity mapping.

Now, given  a non-empty closed convex set, say $\mathcal{C}\subset\R^m$, and a convex function $\Theta:\R^m\rightarrow\R$, such that $\mathcal{C} \cap \levs\Theta\neq\emptyset$ it holds that \cite{yukawaquadtratic}:
\begin{equation}
\forall \wo\in\R^m,\forall \what\in \mathcal{C}\cap\levs\Theta:  \frac{2-\lambda}{2} \Vert \wo-P_{\mathcal{C}}T_{\Theta,\lambda}^{(\G)}(\wo) \Vert^2_{\bm{G}}\leq \Vert\wo-\what\Vert^2_{\bm{G}}-\Vert P_{\mathcal{C}}T_{\Theta,\lambda}^{(\G)}(\wo)-\what\Vert^2_{\bm{G}}.
\label{quasinonexp}
\end{equation}
Following similar steps as in \cite{kopsinis}, it can be proved that $\forall n\geq z_0,\ \phio_{k,n}\in\levs\Theta_{k,n}\Leftrightarrow\I=\emptyset$  and $\forall n\geq z_0, \phio_{k,n}\notin\levs\Theta_{k,n}\Leftrightarrow\I\neq\emptyset$. Moreover, $\levs\Theta_{k,n}=\bigcap_{j\in\I}S_{k,j}\supset\Omega_n\supset\Omega$.  Recall the definition of the relaxed projection mapping;  it can be readily seen that $\wo_{k,n+1}=\Prb T_{\Theta_{k,n},\lambda_{k,n}}^{(\G)}(\phio_{k,n})$. Exploiting this fact,  under Assumptions (a), (b), and (\ref{quasinonexp}) we have that
\begin{align}
\forall n\geq z_0,\forall k\in \N,&\forall \what\in \Omega: \nonumber\\ 0 &\leq  \frac{2-\lambda_{k,n}}{2} \Vert \phio_{k,n}-\wo_{k,n+1} \Vert^2_{\bm{G}} = \frac{2-\lambda_{k,n}}{2} \Vert \phio_{k,n}-\Prb T_{\Theta_{k,n},\lambda_{k,n}}^{(\G)}(\phio_{k,n}) \Vert^2_{\bm{G}}\nonumber \\ &\leq \Vert\phio_{k,n}-\what\Vert^2_{\bm{G}}-\Vert \Prb T_{\Theta_{k,n},\lambda_{k,n}}^{(\G)}(\phio_{k,n})-\what\Vert^2_{\bm{G}}.
\label{quasi2}
\end{align}
Recalling the definitions $\bm{\underline{h}}_{n}=[\wo_{1,n}^T,\ldots,\wo_{K,n}^T]^T\in\R^{Km}$, $\bm{P}_n\bm{\underline{h}}_n=[\phio_{1,n}^T,\ldots,\phio_{K,n}^T]^T\in\R^{Km}$, and (\ref{quasi2}), we have
\begin{align}
\forall n\geq z_0, &\forall \underline{\what}\in \mathfrak{C}: \nonumber\\ 0 &\leq  \min_k\left\lbrace\frac{2-\lambda_{k,n}}{2}\right\rbrace \Vert \bm{P}_n\bm{\underline{h}}_n-\bm{\underline{h}}_{n+1} \Vert^2_{\Ghat} \nonumber \\ &\leq \Vert\bm{P}_n\bm{\underline{h}}_n-\underline{\what}\Vert^2_{\Ghat}-\Vert \bm{\underline{h}}_{n+1}-\underline{\what}\Vert^2_{\Ghat}.
\label{eqdistance}
\end{align}
Nevertheless, from Claim 2, the previous inequality can be rewritten
\begin{equation*} 0 \leq \Vert \bm{P}_n\bm{\underline{h}}_n-\underline{\what} \Vert^2_{\Ghat}-\Vert \bm{\underline{h}}_{n+1}-\underline{\what}\Vert^2_{\Ghat} \leq \Vert\bm{\underline{h}}_n-\underline{\what}\Vert^2_{\Ghat}-\Vert \bm{\underline{h}}_{n+1}-\underline{\what}\Vert^2_{\Ghat}.
\end{equation*}
Hence,\begin{equation}\forall n\geq z_0, \forall \underline{\what}\in \mathfrak{C}:\ \Vert \bm{\underline{h}}_{n+1}-\underline{\what}\Vert^2_{\Ghat} \leq \Vert\bm{\underline{h}}_n-\underline{\what}\Vert^2_{\Ghat}, \label{monotone}\end{equation}
 which completes our proof.\QED
\subsection{Asymptotic optimality}
%The subdifferential of a convex function $\Theta:\R^m\rightarrow\R$ with respect to the $\bm{G}$ metric, is the set of all subgradients, i.e., $\partial_{(\G)}\Theta(\wo):=\lbrace\bm{s}\in\R^m:\langle\wo-\bm{x},\bm{s}\rangle_{\G}+\Theta(\bm{w})\leq\Theta(\bm{x}),\forall \bm{x}\in\R^m\rbrace$. 
A well known property of the projection operator (see  for example \cite{yukawaquadtratic}), is the non-expansivity, i.e., given a non-empty set $\mathcal{C}$, $\Vert P^{(\G)}_{\mathcal{C}}(\bm{h}_1)-P^{(\G)}_{\mathcal{C}}(\bm{h}_2) \Vert_{\G}\leq\Vert\wo_1-\wo_2\Vert_{\G},\forall \wo_1,\wo_2\in\R^m$.
Recall the definition of the algorithm given in  (\ref{theta}). Then, $\forall k\in\N,\forall n\geq z_0,\forall \what\in\Omega$, we have
\begin{align}\Vert\wo_{k,n+1}-\what\Vert_{\G} &= \left\Vert P^{(\bm{G})}_{B_{\ell_1}[\psio_n,\rho]}\left( \phio_{k,n}-\lambda_{k,n}\dfrac{\Theta_{k,n}(\phio_{k,n})}{\Vert\Theta_{k,n}'(\phio_{k,n})\Vert_{\bm{G}}^2} \Theta_{k,n}'(\phio_{k,n}) \right)-\what\right\Vert_{\G} \nonumber \\ &=\left\Vert P^{(\bm{G})}_{B_{\ell_1}[\psio_n,\rho]}\left( \phio_{k,n}-\lambda_{k,n}\dfrac{\Theta_{k,n}(\phio_{k,n})}{\Vert\Theta_{k,n}'(\phio_{k,n})\Vert_{\bm{G}}^2} \Theta_{k,n}'(\phio_{k,n}) \right)-\Prb(\what)\right\Vert_{\G} \nonumber \\  &\leq \left\Vert \phio_{k,n}-\lambda_{k,n}\dfrac{\Theta_{k,n}(\phio_{k,n})}{\Vert\Theta_{k,n}'(\phio_{k,n})\Vert_{\bm{G}}^2} \Theta_{k,n}'(\phio_{k,n}) -\what\right\Vert_{\G},\label{nodealg}\end{align}
where the equality in the second line holds  since, by definition, $\what\in\Omega\subset B_{\ell_1}[\psio_n,\rho]$ and the inequality, from the non-expansivity of the projection operator. Assuming that  $\Ths\neq\bm{0},\forall k\in \N$, and rewriting (\ref{nodealg}) for all the nodes of the network we have
\begin{align}
 \left\Vert \underline{\wo}_{n+1}-\underline{\what}\right\Vert_{\Ghat}^{2}  &\leq \nonumber  \left\Vert \left[\begin{array}{c}
\phio_{1,n}-\lambda_{1,n}\frac{\Theta_{1,n}(\phio_{1,n})}{\left\Vert \Theta_{1,n}'(\phio_{1,n})\right\Vert_{\G} ^{2}}\Theta_{1,n}'(\phio_{1,n})\\
\vdots\\
\phio_{K,n}-\lambda_{K,n}\frac{\Theta_{K,n}(\phio_{K,n})}{\left\Vert \Theta_{K,n}'(\phio_{K,n})\right\Vert_{\G} ^{2}}\Theta_{K,n}'(\phio_{K,n})\end{array}\right]  -\underline{\what}\right\Vert_{\Ghat} ^{2}\nonumber \\
&= \left\Vert  \left[ \begin{array}{c}\phio_{1,n}- \what \\ 
\vdots \\
\phio_{K,n}- \what
 \end{array} \right] \right\Vert_{\Ghat}^{2}+  \sum_{k\in\N}\lambda_{k,n}^{2}\frac{\left(\Theta_{k,n}(\phio_{k,n})\right)^{2}}{\left\Vert \Theta_{k,n}'(\phio_{k,n})\right\Vert_{\G}^{2}}\nonumber \\
&- 2\sum_{k\in\N}\lambda_{k,n}\frac{\Theta_{k,n}(\phio_{k,n})\langle\Theta_{k,n}'(\phio_{k,n}),\left(\phio_{k,n}-\what\right)\rangle_{\G}}{\left\Vert \Theta_{k,n}'(\phio_{k,n})\right\Vert_{\G}^{2}}. \label{eq14} \end{align}
Nevertheless, 
\begin{equation}\left\Vert  \left[ \begin{array}{c}\phio_{1,n}- \what \\ 
\vdots \\
\phio_{K,n}- \what
 \end{array} \right]\right\Vert_{\Ghat}=\Vert\bm{P}_n\bm{\underline{h}}_n-\underline{\what}\Vert_{\Ghat}\leq\Vert\bm{\underline{h}}_n-\underline{\what}\Vert_{\Ghat}.\label{consterm}\end{equation}
From the definition of the subgradient, we have 
\begin{equation}
\langle\Theta_{k,n}'(\phio_{k,n}),\left(\phio_{k,n}-\what\right)\rangle_{\G}\geq \Theta_{k,n}(\phio_{k,n}) -  \Theta_{k,n}(\what)=  \Theta'_{k,n}(\phio_{k,n}),
\label{subeq1}
\end{equation}
where the last equation, holds due to the fact that $\what\in\Omega \Leftrightarrow \Theta'_{k,n}(\what)=\bm{0}$.
Taking (\ref{consterm}) and (\ref{subeq1}) into consideration,  we obtain
\begin{align}\left\Vert \underline{\bm{h}}_{n+1}-\underline{\what}\right\Vert_{\Ghat}^{2} \leq
\left\Vert \underline{\bm{h}}_{n}-\underline{\what}\right\Vert_{\Ghat}^{2} -\sum_{k\in\N}\lambda_{k,n}(2-\lambda_{k,n})\frac{\Th}{\Vert\Ths\Vert^2_{\G}}.
\label{thet}
 \end{align}
Here, notice that the sequence $\left\Vert \underline{\bm{h}}_{n}-\underline{\what}\right\Vert_{\Ghat}$ is bounded and monotone decreasing, hence it converges. The latter fact implies that
\begin{equation}
\lim_{n\rightarrow\infty}\left(\left\Vert \underline{\bm{h}}_{n}-\underline{\what}\right\Vert_{\Ghat}-\left\Vert \underline{\bm{h}}_{n+1}-\underline{\what}\right\Vert_{\Ghat}\right)=0.
\label{seqlimit}
\end{equation}
Under Assumption (c), (\ref{thet}) can be rewritten
\begin{align}\sum_{k\in\N}\varepsilon_1^2\frac{\Th}{\Vert\Ths\Vert^2_{\G}} \leq \sum_{k\in\N}\lambda_{k,n}(2-\lambda_{k,n})\frac{\Th}{\Vert\Ths\Vert^2_{\G}} \leq \left\Vert \underline{\bm{h}}_{n}-\underline{\what}\right\Vert_{\Ghat}^{2} -
\left\Vert \underline{\bm{h}}_{n+1}-\underline{\what}\right\Vert_{\Ghat}^{2}.
\label{thet2}
 \end{align}
Taking limits in (\ref{thet2}) and recalling (\ref{seqlimit})  we have that \begin{equation*}\lim_{n\rightarrow \infty}\frac{\Th}{\Vert\Ths\Vert^2_{\G}}=0,\quad \forall k\in\N.\end{equation*}
If we follow similar steps as in \cite{kopsinis}, it can be verified that $\forall n\in\Z,\forall k\in\N,\forall \wo\in\R^m:\Vert\Theta'_{k,n}(\bm{h})\Vert_{\G}\leq 1$. So, if $\Ths\neq\bm{0}$
\begin{equation}
\Th\leq \frac{\Th}{\Vert\Ths\Vert^2_{\G}}\rightarrow 0,\ n\rightarrow\infty.
\label{theta1}
\end{equation}
Obviously, recalling the previous discussion, $\Ths=\bm{0}\Leftrightarrow \Th=0, \forall n\geq z_0$. Combining this fact together with (\ref{theta1}), we have that 
\begin{equation}
\forall k\in\N,\ \lim_{n\rightarrow\infty}\Th=0.
\label{limth}
\end{equation}
Now, following similar steps as in  \cite{kopsinis}, it can be shown that there exists $D>0$ such that $L_{k,n}\leq D,\forall k\in\N, \forall n\in\Z$.
From the definition of $\Theta_{k,n}$, and under Assumption (d), we have $\forall k\in\N$
\begin{align*}
\frac{D}{\tilde{\omega}_k}\Th &\geq \frac{D}{\tilde{\omega}_k} \sum_{j\in\J}\omega_{k,j}\frac{{\dif^2(\phio_{k,n},S_{k,j})}}{L_{k,n}} \nonumber \\
&\geq \frac{D}{\tilde{\omega}_k} \frac{\tilde{\omega}_k}{D} \sum_{j\in\J}{{\dif^2(\phio_{k,n},S_{k,j})}} \nonumber \\
&\geq \max \lbrace\dif^2(\phio_{k,n},S_{k,j}):j\in\J \rbrace.
\end{align*}
Taking limits in the previous inequality, we obtain that 
\begin{equation}
\lim_{n\rightarrow\infty} \max \lbrace\dif(\phio_{k,n},S_{k,j}):j\in\J \rbrace =0.
\label{phidist}
\end{equation}
Combining (\ref{eqdistance}) with the result of Claim 2, we have
\begin{align}
\forall n\geq z_0, &\forall \underline{\what}\in \mathfrak{C}: \nonumber\\ 0 &\leq  \min_k\left\lbrace\frac{2-\lambda_{k,n}}{2}\right\rbrace \Vert \bm{P}_n\underline{\wo}_n-\bm{\underline{h}}_{n+1} \Vert^2_{\Ghat} \nonumber \\ &\leq \Vert \bm{\underline{h}}_n-\underline{\what}\Vert^2_{\Ghat}-\Vert \bm{\underline{h}}_{n+1}-\underline{\what}\Vert^2_{\Ghat}.
\label{eqdist2}
\end{align}
Taking limits in (\ref{eqdist2}) and recalling (\ref{seqlimit}) gives us
\begin{align}
\lim_{n\rightarrow\infty} \Vert \bm{P}_n\bm{\underline{h}}_n-\bm{\underline{h}}_{n+1} \Vert_{\Ghat}^2=0 \Leftrightarrow \lim_{n\rightarrow\infty} \sum_{k\in\N}\Vert\phio_{k,n}-\wo_{k,n+1}\Vert_{\G}^2=0.
\label{consineq}
\end{align}
 \bibliographystyle{IEEEtran}
 Fix an arbitrary point $\bm{v}\in S_{k,j},\forall k\in\N, \forall j\in\J$. Then from the triangle inequality we have
 \begin{align}
\Vert \wo_{k,n+1}-\bm{v}\Vert_{\G} &\leq  \Vert \wo_{k,n+1}-\phio_{k,n}\Vert_{\G}+ \Vert \phio_{k,n}-\bm{v}\Vert_{\G}  \Rightarrow \nonumber \\ 
\inf_{\bm{v}\in S_{k,j}} \Vert \wo_{k,n+1}-\bm{v}\Vert_{\G} &\leq  \Vert \wo_{k,n+1}-\phio_{k,n}\Vert_{\G}+ \inf_{\bm{v}\in S_{k,j}} \Vert \phio_{k,n}-\bm{v}\Vert_{\G}  \Rightarrow \nonumber \\ \dif(\wo_{k,n+1},S_{k,j}) &\leq  \Vert \wo_{k,n+1}-\phio_{k,n}\Vert_{\G}+  \dif(\phio_{k,n},S_{k,j})
\label{disttri} 
 \end{align}
 If we take limits in (\ref{disttri}), from (\ref{phidist}) and (\ref{consineq}), it can be seen that 
 \begin{align}
 \lim_{n\rightarrow\infty} \dif(\wo_{k,n+1},S_{k,j})=0, \forall k\in\N, \forall j\in\J\Leftrightarrow 
  \lim_{n\rightarrow\infty} \sum_{j\in\J} \dif(\wo_{k,n+1},S_{k,j})=0, \ \forall k\in \N.
\label{lim1}
 \end{align}
The definitions of the distance function and the projection operator, yield
\begin{align}
\mathrm{d}(\wo_{k,n+1},S_{k,j})&=\Vert \wo_{k,n+1}-P_{S_{k,j}}(\wo_{k,n+1})\Vert \nonumber\\
                            &\leq \Vert \wo_{k,n+1}-P^{(\G)}_{S_{k,j}}(\wo_{k,n+1})\Vert.
\label{lim2}
\end{align}  
 Nevertheless, the Rayleigh-Ritz theorem implies  \cite{horn2005matrix} $\forall \wo\in\R^m: \Vert\wo\Vert \leq \tau_{\min}^{-\frac{1}{2}}\Vert\wo\Vert_{\G}$, where $\tau_{\min}$ is the smallest eigenvalue of $\G$. Combining this fact as well as (\ref{lim2}) we obtain
\begin{align}
\mathrm{d}(\wo_{k,n+1},S_{k,j})&\leq\Vert \wo_{k,n+1}-P^{(\G)}_{S_{k,j}}(\wo_{k,n+1})\Vert \nonumber\\
                            &\leq\tau_{min}^{-\frac{1}{2}} \Vert \wo_{k,n+1}-P^{(\G)}_{S_{k,j}}(\wo_{k,n+1})\Vert_{\G}\rightarrow 0,n\rightarrow \infty, \forall k\in\N,
\end{align}
where the limit holds from (\ref{lim1}). From the previous, it is not difficult to obtain that
\begin{align*}
\lim_{n\rightarrow\infty}\max\lbrace\mathrm{d}(\wo_{k,n+1},S_{k,j}):j\in\J\rbrace=0,
\end{align*}
 which completes our proof. \QED
\subsection{Asymptotic Consensus}
 In \cite{cavalcante2009adaptive} it has been proved, that the algorithmic scheme achieves asymptotic consensus, i.e., $\Vert\wo_{k,n}-\wo_{l,n}\Vert\rightarrow 0,\ n\rightarrow \infty, \forall k,l\in\N$ if and only if
\begin{equation}
\lim_{n\rightarrow\infty}\Vert\bm{\underline{h}}_n-P_{\mathcal{O}}(\bm{\underline{h}}_n)\Vert={0}.
\label{ascons}
\end{equation}
Let Assumptions (a), (b), (c), (d), hold true. We define the following quantity 
\begin{equation}
\underline{\bm{\epsilon}}_n:=\bm{\underline{h}}_{n+1}-\bm{P}_n\bm{\underline{h}}_n.
\label{cons21}
\end{equation}
 Obviously from (\ref{consineq})
\begin{equation}
\underset{n\rightarrow\infty}{\lim}\eb_n=\bm{0}
\label{K17}
\end{equation}
Now, if we rearrange the terms in (\ref{cons21}) and if we iterate the resulting equation, we have:
\begin{align}
\bm{\underline{h}}_{n+1} &=\bm{P}_n\bm{\underline{h}}_n+\eb_n \nonumber \\ &= \bm{P}_n\bm{P}_{n-1}\underline{\wo}_{n-1} + \bm{P}_n\eb_{n-1}+\eb_n=\ldots \nonumber  \\ &=\prod_{i=1}^n \bm{P}_i\underline{\wo}_0 +   \sum_{j=1}^{n}\prod_{l=0}^{n-j} \bm{P}_{n-l}\eb_{j-1} + \eb_{n} \nonumber 
\end{align}
If we left-multiply the previous equation by $(\bm{I}_{Km}-\bm{B}\bm{B}^{T})$, where $\bm{I}_{Km}$ is the ${Km}\times {Km}$  identity matrix, and  follow similar steps as in \cite[Lemma 2]{cavalcante2009adaptive} it can be verified  that
$\underset{n \rightarrow \infty}{\lim} \Vert\left(\bm{I}_{Km}-\bm{B}\bm{B}^{T}\right)\bm{\underline{h}}_{n+1}\Vert=0$
which completes our proof.\QED

\subsection{Strong Convergence}
We will prove, that under assumptions (a), (b), (c), (d), (e), $\lim_{n\rightarrow\infty}\bm{\underline{h}}_n=\hat{\underline{\bm{h}}}_*,\hat{\underline{\bm{h}}}_*\in\mathcal{O}$.
Recall that the projection operator, of an arbitrary vector $\underline{\bm{h}}\in\mathbb{R}^{Km}$ onto the consensus subspace equals to $P_{\mathcal{O}}(\underline{\bm{h}})=\bm{B}\bm{B}^T\underline{\bm{h}},\forall \underline{\bm{h}}\in\mathbb{R}^{Km}$. Taking into consideration Assumption (e) together with (\ref{monotone}), from \cite[Lemma 1]{yamada2004adaptive} we have that there exists $\hat{\underline{\bm{h}}}_*\in\mathcal{O}$ such that 
\begin{equation}
\lim_{n\rightarrow\infty}P_{\mathcal{O}}(\bm{\underline{h}}_n)=\hat{\underline{\bm{h}}}_*.
\label{sub_conv}
\end{equation}
Now, exploiting the triangle inequality we have that
\begin{equation}
\Vert\bm{\underline{h}}_n-\hat{\underline{\bm{h}}}_*\Vert \leq \Vert\bm{\underline{h}}_n-P_{\mathcal{O}}(\bm{\underline{h}}_n)\Vert + \Vert\hat{\underline{\bm{h}}}_*-P_{\mathcal{O}}(\bm{\underline{h}}_n)\Vert \rightarrow 0, \ n\rightarrow\infty,
\label{str_con}
\end{equation}
where this limit holds from (\ref{ascons}) and (\ref{sub_conv}). The proof is complete since (\ref{str_con}) implies that
$\lim_{n\rightarrow\infty}\bm{\underline{h}}_n=\hat{\underline{\bm{h}}}_*$.\QED

\bibliographystyle{IEEEtran}
 \bibliography{ref4}
 \end{document}